\documentclass[
    ,final            
  ]
  {aipproc}

\layoutstyle{6x9}

\usepackage{amsmath}
\begin{document}

\title{The Penn State - Toru\'n Planet Search: Target Characteristics and Recent Results}


\author{Pawe\l{} Zieli\'nski}{
  address={Toru\'n Centre for Astronomy, Nicolaus Copernicus University, Gagarina 11, 87-100 Toru\'n, Poland}
}

\author{Andrzej Niedzielski}{
  address={Toru\'n Centre for Astronomy, Nicolaus Copernicus University, Gagarina 11, 87-100 Toru\'n, Poland}
}

\author{Aleksander Wolszczan}{
  address={Department for Astronomy and Astrophysics, Pennsylvania State University, 525 Davey Laboratory, University Park, PA 16802, USA}
  ,altaddress={Center for Exoplanets and Habitable Worlds, Pennsylvania State University, 525 Davey Laboratory, University Park, PA 16802, USA} 
}

\author{Grzegorz Nowak}{
  address={Toru\'n Centre for Astronomy, Nicolaus Copernicus University, Gagarina 11, 87-100 Toru\'n, Poland}
}

\author{Monika Adam\'ow}{
  address={Toru\'n Centre for Astronomy, Nicolaus Copernicus University, Gagarina 11, 87-100 Toru\'n, Poland}
}

\author{Sara Gettel}{
  address={Department for Astronomy and Astrophysics, Pennsylvania State University, 525 Davey Laboratory, University Park, PA 16802, USA}
  ,altaddress={Center for Exoplanets and Habitable Worlds, Pennsylvania State University, 525 Davey Laboratory, University Park, PA 16802, USA} 
}

\begin{abstract}
More than 450 stars hosting planets are known today but only approximately 30 planetary systems were discovered around stars beyond the Main Sequence. The Penn State-Toru\'n Planet Search, putting an emphasis on extending studies of planetary system formation and evolution to intermediate-mass stars, is oriented towards the discoveries of substellar-mass companions to a large sample of evolved stars using high-precision radial velocity technique. We present the recent status of our survey and detailed characteristic for $\sim$350 late type giant stars, i.e.\ the new results of radial velocity analysis and stellar fundamental parameters obtained with extensive spectroscopic method. Moreover, in the future we will make an attempt to perform the statistical study of our sample and searching the correlations between the existence of substellar objects and stellar atmospheric parameters according to previous works which investigated the planetary companion impact on the evolution of the host stars. 
\end{abstract}

\maketitle


\section{Introduction}
The search for planets around giants represent an essential complement to ``traditional'' surveys, because they furnish information about properties of planetary systems around stars that are the descendants of the A--F main sequence (MS) stars with masses as high as $\sim5~{\rm M}_{\odot}$. As the stars evolve off the MS, their effective temperatures and rotation rates decrease to the point that their radial velocity (RV) variations can be measured with a few 
m\,s$^{-1}$ precision. This offers an excellent opportunity to improve our understanding of the population of planets around stars that are significantly more massive than the Sun. Without such surveys it would be difficult to construct a broad, integrated picture of planet formation and evolution.

Since 2001, about 30 such objects have been identified, including our five published detections with the Hobby-Eberly Telescope (HET, \citealt{N2007,N2009a, N2009b}). Our work has produced the tightest orbit of a planet orbiting a K-type giant identified so far (0.6 AU) and the first convincing evidence for a multi-planet system around such a star \citep{N2009a}. Our recent discoveries \citep{N2009b} have also identified new multi-planet systems, including a very intriguing one of two brown dwarf-mass bodies orbiting a $2.8~{\rm M}_{\odot}$, K2 giant. This particular detection challenges the standard interpretation of the so-called brown dwarf desert known to exist in the case of solar-mass stars. Along with the discoveries supplied by other groups our work has substantially added to the emerging evidence that stellar mass positively correlates with masses of substellar companions, all the way from red dwarfs to intermediate-mass stars. 

In this paper we present the current status and forthcoming results from the Penn State-Toru\'n Planet Search (PTPS) around G--K giants with the HET since 2004.

\section{Targets sample}
High-precision radial velocity determination for MS stars more massive than $\sim1.3~{\rm M}_{\odot}$ are not possible due to small number of spectral lines present in the spectra of these stars and their significant rotational broadening. However, it is possible to obtain precise RVs for intermediate-mass stars after they have left the MS and become subgiants or giants cool enough to develop spectra that are rich in narrow lines. Among the various planet detection methods in the case of stars more massive than $\sim1~{\rm M}_{\odot}$ RV searches provide the most efficient way to do it, while for stars more massive than $\sim2~{\rm M}_{\odot}$ they are practically the only way.

The whole PTPS targets sample of nearly 1000 late-type stars observed with this survey (Fig.~1) comprises about 350 giants from the giant clump, about 350 giants and subgiants, and roughly 250 evolved stars in the upper envelope of the MS (aging dwarfs). 
In terms of using the RV technique, this star sample is limited to 0.4$^{{\rm m}}$ $\le(B-V)\le$ 1.2$^{{\rm m}}$ and concentrated around 1.5$^{{\rm m}}$ above the MS. Moreover, all PTPS targets are uniformly distributed in right ascension and declination on the sky. The magnitude upper limit of the stars in the $V$ band is 11$^{{\rm m}}$ and usually (> 66~\%) the program stars are fainter than 8$^{{\rm m}}$.

\begin{figure}
\includegraphics[height=0.8\textwidth,angle=270]{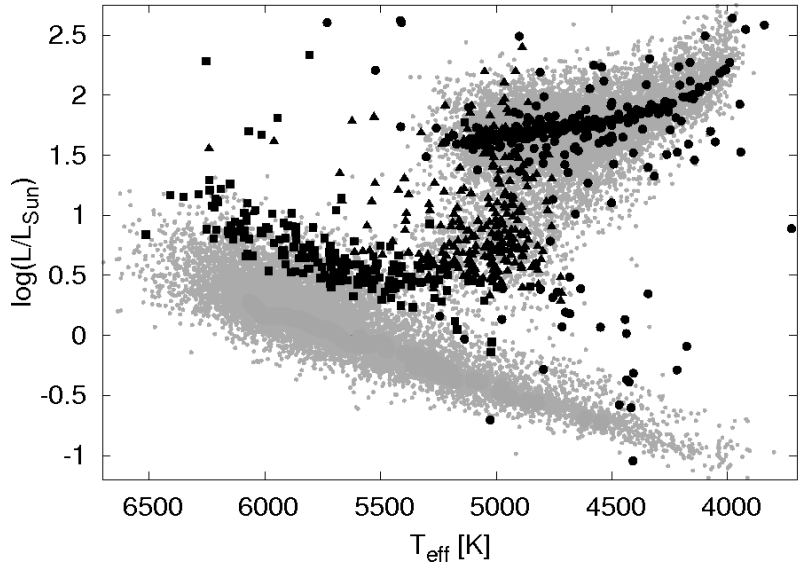}
\caption{The PTPS survey stars: clump giants (circles) contains stars of various masses over a range of evolutionary tracks; intermediate-mass dwarfs (squares) which are about to leave the MS; giants and subgiants (triangles). Background stars (grey points) are from The Tycho-2 Catalogue and are not included in the PTPS.}
\end{figure}

\section{Observations and data reduction}
Observations for the PTPS have been obtained with the $9.2\,{\rm m}$ HET \citep{R1998} equipped with the High Resolution Spectrograph (HRS, R=60\,000, \citealt{Tu1998}) in the queue scheduled mode \citep{Sh2007}. The spectrograph was fed with a 2~arcsec fiber. The spectra consisted of 46 echelle orders recorded on the ``blue'' CCD chip (407--592 nm) and 24 orders on the ``red'' chip (602--784 nm). A typical signal-to-noise ratio was 200--250 per resolution element. The basic data reduction was performed in a standard manner using the IRAF\footnote{IRAF is distributed by the National Optical Astronomy Observatories and operated by the Association of Universities for Research in Astronomy, Inc., under cooperative agreement with the National Science Foundation.} tasks and scripts. 

The observing strategy was focused on long-term variations in RVs which were measured using the standard I$_{2}$ cell calibration technique \citep{B1996}. Details of our survey, the observing procedure and data analysis, have been described elsewhere \citep{N2007,NW2008}.

\section{Stellar characteristics and results}
Late type giants are known to exhibit various types of intrinsic variability induced by pulsations and/or surface inhomogeneity connected with rotation. These effects have to be ruled out before substellar companion interpretation can be considered. Therefore, proper interpretation of results obtained from RV survey of G--K giants requires detailed knowledge of their physical parameters.  

\subsection{Atmospheric parameter determination}
An atmospheric model of the star is characterized by effective temperature $T_{\rm eff}$, surface gravity $\log\,{g}$, microturbulence velocity $v_{\rm t}$, and metallicity [Fe/H]. To obtain these parameters from high resolution spectra, we used a purely spectroscopic method developed by \citet{T2002,T2005a,T2005b} which is based on the measurements of equivalent widths (EWs) of iron lines and resulting from the assumption of LTE. We were able to find and measure the EWs of $\sim$300 Fe\,{\sc I} and Fe\,{\sc II} lines in our spectra. To avoid a perturbation of the mean iron abundance (and thus [Fe/H]) dependency on the excitation potential and EWs, we selected the lines not stronger than 150--200 m\AA.

Figure~2 presents the results determined from the method mentioned above and the dependencies between atmospheric parameters ($T_{\rm eff}$, $\log\,{g}$, 
$v_{\rm t}$, and [Fe/H]) in the case of clump giants. The consistency of the solutions is clearly visible and remains in agreement with common features of Red Giant Clump (RGC) stars. The average uncertainties of the following parameters are $\sigma T_{\rm eff} = 14\,{\rm K}$, $\sigma\log\,{(g[{\rm cm\,s}^{-2}])} = 0.05$, $\sigma v_{\rm t} = 0.08\, {\rm km\,s}^{-1}$, and $\sigma$[Fe/H]$ = 0.07$. Only for individual stars these values are slightly higher.

\begin{figure}
\includegraphics[width=0.7\textwidth,angle=270]{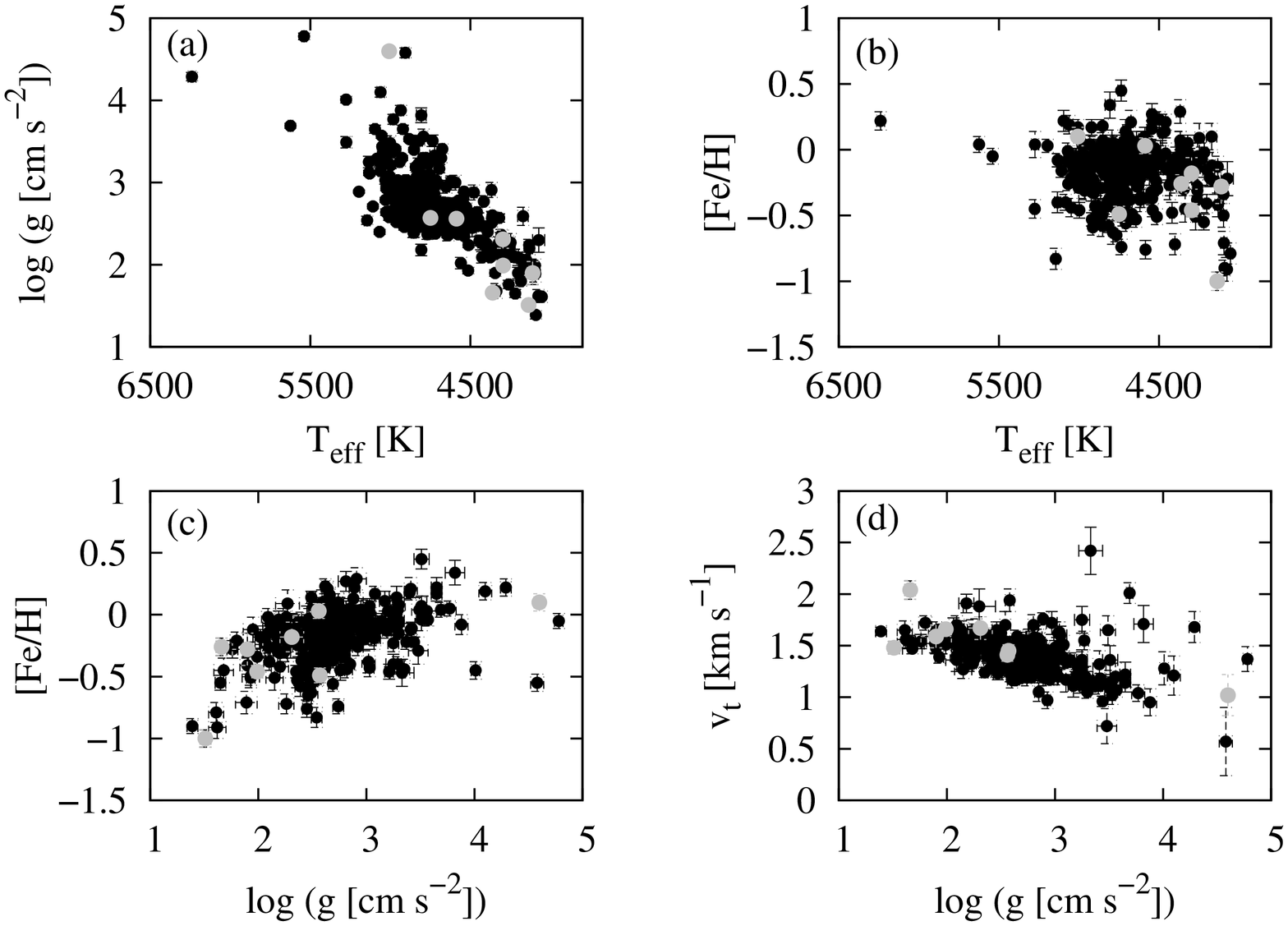}
\caption{Dependencies between four atmospheric parameters for PTPS clump giants:
\mbox{(a) ~ $T_{\rm eff}$ ~versus~ $\log\,{g}$,} ~
\mbox{(b) ~ $T_{\rm eff}$ ~versus~ [Fe/H],} ~
\mbox{(c) ~ $\log\,{g}$ ~versus~ [Fe/H],} ~
 and ~
\mbox{(d) ~ $\log\,{g}$ ~versus~ $v_{\rm t}$.} 
Grey points indicate eight stars with substellar-mass companions discovered by the PTPS.}
\end{figure}

\subsection{Integral parameters determination}
The knowledge of $T_{\rm eff}$, $\log\,{g}$, and [Fe/H] together with Hipparcos/Tycho data allows the estimation of stellar bolometric magnitudes 
M$_{\rm bol}$ and luminosities $\log(L/{\rm L}_{\odot})$. The rest of necessary quantities, such as intrinsic colour $(B-V)_{0}$ and bolometric correction BC, can be determined for each star using empirical calibration taken from \citet{A1999}. To obtain the stellar mass, radius, and age, i.e.\ the parameters constraining the stellar evolution, the position of the each single star should be compared with appropriate evolutionary tracks and isochrones on the H-R diagram \citep{G2000,S2000}, although the accurate estimation of these parameters is made very difficult due to the complex nature of the RGC region in the H-R diagram.

Figures~3 and 4 show the distributions of masses and radii estimated from the general definitions based on $T_{\rm eff}$, $\log\,{g}$, and $\log(L/{\rm L}_{\odot})$. It is worthwhile noting that these estimates do not take into account the stellar composition, which is a very important factor influencing the stellar evolution. Nevertheless, both distributions under consideration confirm that we study the objects in the range 1--3 ${\rm M}_{\odot}$ and 8--13 ${\rm R}_{\odot}$. 

The typical systematical errors are $\sigma\log(L/{\rm L}_{\odot}) = 0.4$, $\sigma M = 0.3\, {\rm M}_{\odot}$, and $\sigma R = 0.8\, {\rm R}_{\odot}$. However, we should mention that for many cases the uncertainties are much higher because of the moderate quality of stellar photometry available in the astronomical databases and/or high uncertainties in Hipparcos parallaxes.

\begin{figure}
\includegraphics[width=0.54\textwidth]{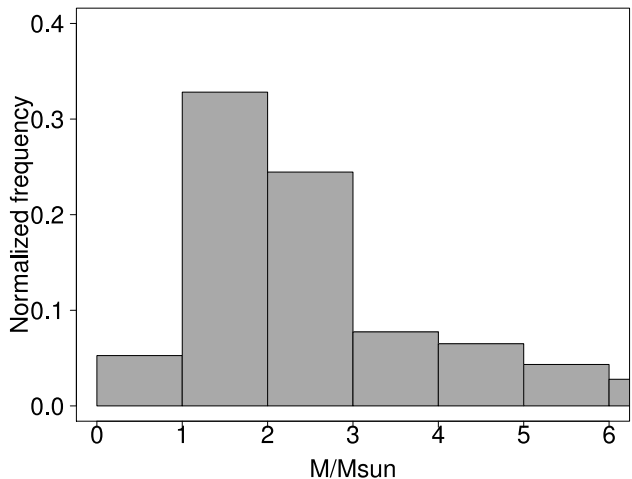}
\caption{Masses distribution of the PTPS clump giants. Most of the stars fall in the 1--3 ${\rm M}_{\odot}$ range.}
\end{figure}

\begin{figure}
\includegraphics[width=0.54\textwidth]{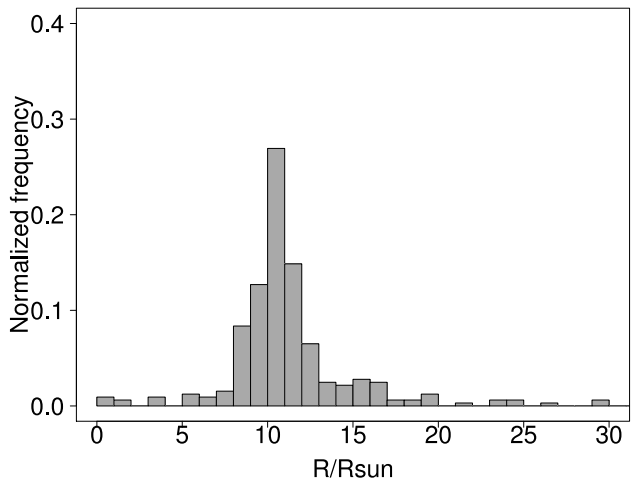}
\caption{Radii distribution of the PTPS clump giants. Most of the stars fall in the 8--13 ${\rm R}_{\odot}$ range.}
\end{figure}

\section{Summary and conclusions}
Since 2004, conclusive multi-epoch observations have been gathered for about 800 stars. The large number of data allow for a preliminary assessment of possible companion candidates, and, depending on the criteria adopted, the substellar-mass companion frequency may reach up to 30$\%$ in the case of the whole PTPS sample.

\begin{figure}
\includegraphics[width=0.6\textwidth]{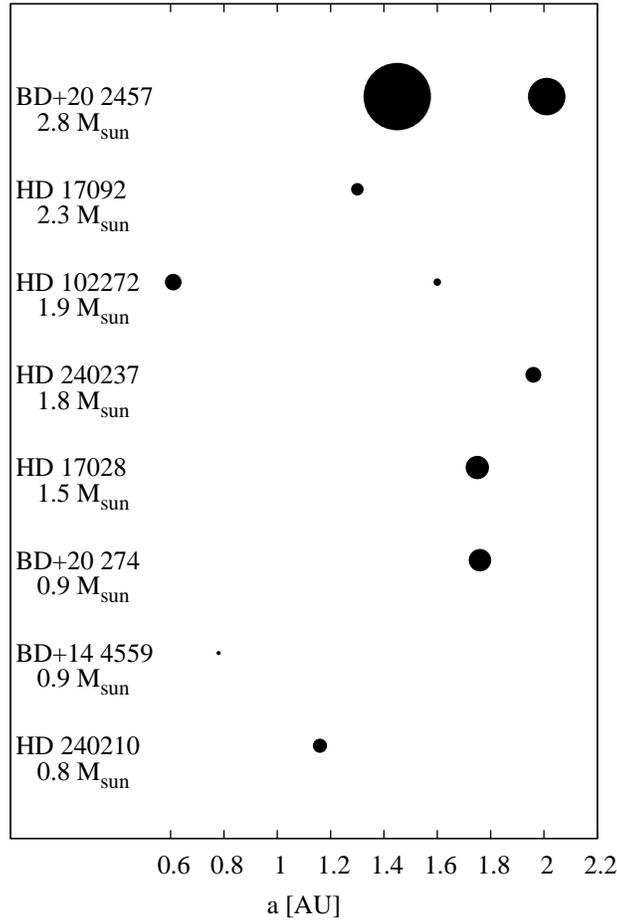}
\caption{The first ten substellar-mass companions discovered by the PTPS in function of the distance from the hosting stars. The symbol sizes are proportional to planet mass.}
\end{figure}

Until today seven stars from PTPS survey have been confirmed to be orbited by planetary systems. The first ten substellar-mass companions discovered by the PTPS are shown in Fig.~5. In addition to the published planet and brown dwarf companions, we also present two new planets around BD+20 274 and HD 240237 (Gettel et al.\ in preparation), and one around HD 17028 (Nowak et al.\ in preparation).

The precise knowledge of stellar parameters is also important in terms of the statistical correlations between planet occurrence frequency, their orbital parameters, and stellar mass or metallicity.  The forthcoming study (Zieli{\'n}ski et al.\ in preparation) shall give a new set of data, and together with previous work \citep{Z2010a,Z2010b}, it will bring more information for statistical analysis and improve the general knowledge concerning the evolution of stars and planetary system formation scenarios.


\begin{theacknowledgments}
We acknowledge the financial support from the Polish Ministry of Science and Higher Education through grants N N203 510938 and N N203 386237. AW acknowledges support from NASA grant NNX09AB36G. GN is a recipient of a graduate stipend of the Chairman of the Polish Academy of Sciences. We thank the HET resident astronomers and telescope operators for support. The Hobby-Eberly Telescope (HET) is a joint project of the University of Texas at Austin, the Pennsylvania State University, Stanford University, Ludwig-Maximilians-Universit\"at M\"unchen, and Georg-August-Universit\"at G\"ottingen. The HET is named in honor of its principal benefactors, William P.\ Hobby and Robert E.\ Eberly. The Center for Exoplanets and Habitable Worlds is supported by the Pennsylvania State University, the Eberly College of Science, and the Pennsylvania Space Grant Consortium.
\end{theacknowledgments}

\bibliographystyle{aipproc}   


\end{document}